# Superconductivity with two-fold symmetry in topological superconductor $Sr_xBi_2Se_3$

Guan Du[1], Yufeng Li[1], J. Schneeloch[2], R. D. Zhong[2], Genda Gu[2], Huan Yang[1,3] and Hai-Hu Wen[1,3*]

[1] National Laboratory of Solid State Microstructures and Department of Physics, Nanjing University, Nanjing 210093, China
[2] Condensed Matter Physics and Materials Science Department, Brookhaven National Laboratory, Upton, New York 11973, US
[3] Collaborative Innovation Center of Advanced Microstructures, Nanjing University, Nanjing, 210093, China

**Topological superconductivity is the quantum condensate of paired electrons with an odd parity of the pairing function. One of the candidates is the triplet pairing superconductor derived from topological insulator $Bi_2Se_3$ by chemical doping. Theoretically it was predicted that a two-fold superconducting order parameter may exist in this kind of bulk topological superconductor. Earlier experiments by nuclear magnetic resonance and angle-resolved magnetocaloric measurements seem to support this picture. By using a Corbino-shape like electrode configuration, we measure the *c*-axis resistivity without the influence of the vortex motion and find clear evidence of two-fold superconductivity in the recently discovered topological superconductor $Sr_xBi_2Se_3$. The Laue diffraction measurements on these samples show that the maximum gap direction is either parallel or perpendicular to the main crystallographic axis.**

Topological material has already become one of the hottest topics in condensed matter physics. As an important member of the topological family, topological superconductors (TSCs) have drawn great attention recently[1]. TSCs are novel quantum phases characterized by topologically nontrivial Cooper pairing orders. The unconventional superconductivity in TSCs cannot be connected to a topological trivial phase adiabatically without closing the superconducting gap. As a consequence, a TSC is guaranteed to possess robust gapless excitations on the boundary (or surface). Because of the particle-hole symmetry of superconducting states, the excitation at zero energy is composed of same weight of electrons and holes. Such a zero energy excitation satisfies the requirement of Majorana fermions whose antiparticle is identical to itself[1-3]. For the three dimensional TSC, the Majorana zero energy state is itinerant on the surface, and for the two dimensional TSC, the Majorana zero-energy state is trapped in the vortex core[1-3]. The Majorana fermions in TSCs have various exotic phenomena and can help to realize topological quantum computation[1-3]. A lot of efforts have been devoted to study the one dimensional[4,5], two dimensional[6-8], and three dimensional[9-20] topological systems, however, the experimental realization and detection of TSCs and Majorana fermions are still open issues.

Among the three dimensional systems, superconductors derived from $Bi_2Se_3$ by chemical doping are intensively studied and are predicted to be candidates of TSCs[11,12]. Point contact measurements have detected zero-bias conductance peaks (ZBCPs) on the surface of $Cu_xBi_2Se_3$ which are interpreted to be signatures of Majorana fermions[9,10]. In contrast, the ZBCPs are absent in scanning tunneling spectroscopy (STS) measurements on the same material [14]. Recently, superconductivity has been discovered in $Sr_xBi_2Se_3$ which is supposed to be a promising candidate of TSC[17-19]. However, the STS studies revealed a full superconducting gap without any abnormal in-gap states [20]. The absence of ZBCPs in the STS studies of $Cu_xBi_2Se_3$, $Sr_xBi_2Se_3$ questions whether these materials are TSCs, or the theoretical understanding on the tunneling spectrum needs to be modified, this makes the issue under intense debates. Recently the NMR studies are conducted on $Cu_xBi_2Se_3$, and the unusual two-fold symmetry of the Knight shift in the hexagonal plane is observed[21]. The angle-dependent specific heat measurements on $Cu_xBi_2Se_3$ also shows similar two-fold symmetry[22]. Both results support the nematic superconductivity in the trigonal $D_{3d}$ crystallographic point group, which may indicate a two-fold nodeless superconducting gap[23].

In this paper, we report the angle-dependent resistivity measurements on high-quality $Sr_xBi_2Se_3$ single crystals with the magnetic field in the hexagonal plane and current flowing along *c*-axis. With this configuration, the vortex contribution to resistivity is always the same for any angles. However, we find that all the angle-dependent resistivity obtained in the superconducting transition region has two-fold symmetry in low temperature and magnetic field regions. The corresponding $H_{c2}(0)$ estimated on the maximum resistivity direction is much smaller than that on the minimum resistivity direction. Our data suggest that the superconducting gap has the two-fold symmetry with gap maximum direction parallel or perpendicular to the main crystallographic axis. Along with the NMR[21] and specific heat studies[22], our results have provided supports to the viewpoint that superconductors derived by doping Sr to $Bi_2Se_3$ materials should have a two-fold order parameter and thus be consistent with the TSCs.

**Results**

**Superconducting properties.** All the measurements in the work are performed on 4 different $Sr_xBi_2Se_3$ single crystals. All samples show rather sharp superconducting transitions. Taking sample 1 for example, the temperature dependent magnetization curves obtained with zero-field-cooling (ZFC) and field-cooling (FC) processes are displayed in Fig. 1a. The superconducting transition temperature is about 2.8 K. The calculated magnetic screening volume is about 322% from the ZFC data at 1.8 K. By considering the demagnetization factor $N_m$ = 0.713 determined from the geometric dimensions of the cuboid specimen, the magnetic screening volume is as high as 92.4% even at 1.8 K, indicating good superconducting property of the sample. The resistance measurements were carried out with the electrode configuration presented as the cartoon like picture shown in the right bottom inset of Fig. 1b. Shown in the top inset of Fig. 1b is the realistic top-view photograph of sample 1 with electrodes. The four-electrode method is performed and electrodes are made on the cleaved surfaces with the Corbino-shape like configuration to make sure that the current flows along the *c*-axis of the sample. The main panel of Fig. 1b displays the resistivity versus temperature of sample 1 measured under zero magnetic field. The superconductivity transition also appears to be very sharp and the transition temperature is 2.78 K obtained from the point where the resistivity equals to the half value of normal state value $\rho_n$. The left bottom cartoon figure shows the magnetic field direction. In this case, even the magnetic field is rotated in the plane of the sample, the current is always perpendicular to the magnetic field and the factor involved by vortex motion is always the same. Fig. 1c shows the Laue diffraction patterns of the cleaved surface of sample 1. Because the selenium terminated surface of the $Sr_xBi_2Se_3$ single crystal has a hexagonal structure, the Laue diffraction patterns display a six-fold symmetry in general. By comparing with the simulated pattern shown in the right part of Fig. 1c, the corresponding lattice structure of the Se terminated surface is deduced and shown in the top of Fig. 1c. To show the relationship between the experimental data of different samples more clearly, we define a Se-Se direction as the zero-degree direction, and $\theta$ is the angle between the magnetic field and the selected Se-Se direction in the hexagonal plane as indicated in the figure. Fig. 1d shows the angle-dependent resistivity of sample 1 at 1.9 K and $\mu_0H$ = 0.5 T. The unambiguous anisotropy is viewed with two obvious peaks. The centers of two peaks locate at 86.3° and 266.3° respectively with the angle difference of 180°, which reveals a clear two-fold symmetry feature. Such a two-fold symmetry feature is more dramatically expressed in the polar coordinates representation as shown in the inset of Fig. 1d.

**Two-fold symmetry of c-axis resistivity in the superconducting transition region.** Figure 2 shows the angle-dependent resistivity data measured at various magnetic fields and temperatures for sample 1, 2, and 3. To avoid the possible asymmetric problem, the data shown here have been taken by averaging the raw data taken with positive and negative magnetic fields. All the data taken on different samples hold the two-fold symmetry in general. We can have a closer look at the angle dependent resistivity profiles versus magnetic fields and temperatures, respectively. The data measured at 1.9 K and different fields shown in Fig. 2a, 2c, and 2e are simply dumbbell-shaped below 1 T. When the magnetic field exceeds 1 T, the profiles of the curves become complex but still hold the two-fold symmetry as the dominant one. Actually at this field, the resistivity here has a value already beyond the 50% of the normal state value (see Fig. 3), suggesting the involvement of many different contributions, such as the anisotropic order parameter, the vortex motion dissipation, the scattering of the unpaired electrons, etc. At 5 T where the samples are in the normal states, the curves are regular circles without any anisotropic components. Talking about temperature dependence, the profiles of resistivity obtained at 0.5 T and different temperatures shown in Fig. 2b, 2d, and 2f preserve simply the dumbbell-shape below 2.2 K, and become complex when temperature exceeds 2.2 K. At 4 K where the superconductivity is totally destroyed, the data are also in the form of regular circles showing a perfect isotropic feature. Above all, the general two-fold symmetric feature is the intrinsic property of the superconducting state. At low fields and low temperatures, the data illustrate the resistive dissipation of pure vortex motion. While since in our configuration, the factor of vortex motion should be similar at all angles, therefore the angle dependence here reflects the very basic information of superconductivity, such as the order parameter or upper critical field. The maximum resistivity directions corresponding to the angle $\theta^{max}$ are defined as the directions of the central of dumbbells at 0.5 T and 1.9 K. For sample 1, 2, and 3, $\theta^{max}$ values are 86.3º, 86.7º, and 3.9º respectively. The directions of minimum resistivity ($\theta^{min}$) are defined as the perpendicular directions of $\theta^{max}$, i.e., 176.3º, 176.7º, and 93.9º for sample 1, 2, and 3 respectively. At higher fields and temperatures near the normal state, the situation becomes very complex because many effects will be involved. For example, we further analyzed the data with complex angular symmetry at 1.5 T with Fourier transformation method, and six-fold as well as four-fold symmetric components are found in the raw data (as shown in Supplementary Note 2). We argue that the six-fold symmetric component may contain the information contributed by the six-fold crystal structure, and the tiny four-fold structure may result from the frequency doubling effect of the two-fold component.

**Anisotropy of the in-plane $H_{c2}$.** We further measured the temperature dependent resistivity at angles $\theta^{max}$ and $\theta^{min}$ which are determined from the angular dependent resistivity at low temperature and magnetic field. The data have been averaged with positive and negative fields to remove the possible Hall components. As shown in Fig. 3, superconductivity can be more easily suppressed by the magnetic field applied at $\theta^{max}$ than that applied at $\theta^{min}$. In order to obtain the upper critical field $H_{c2}$ anisotropy between $\theta^{max}$ and $\theta^{min}$ directions, we record $H_{c2}$ by taking the half of the $\rho_n$ as the criterion for different samples, and the data are shown in Fig. 3c, 3f, and 3i. We then fitted the data by the equation,

$$H_{c2}(T) = H_{c2}(0) \frac{\left[1-\left(\frac{T}{T_c}\right)^2\right]^\alpha}{1+\left(\frac{T}{T_c}\right)^2} \qquad (1)$$

$T_c$ is also determined from the point that equals to 50% value of normal state

resistivity at zero-field for each sample. In above equation we choose the two components (1-$t^2$) and (1+$t^2$) since they are the basic ingredients for constructing the temperature dependence of either the thermodynamic critical field, or the coherence length ξ, or the magnetic penetration depth λ. In the case of Ginzburg-Landau theory, we have $\alpha$ = 1. The fitting parameters for the three samples are illustrated in Supplementary Table 1, and the fitting curves are plotted in c, f, and i as solid lines. The resultant anisotropy of the in-plane $H_{c2}(0)$ is from 1.96 to 2.84 (see Supplementary Table 1), i.e., $H_{c2}(0)$ at $\theta^{min}$ is about twice larger than that at $\theta^{max}$. We also notice that the resistivity curves measured on sample 1 and sample 2 have "peak" structures near the onset of superconducting transitions. We suggest that this "peak effect" of resistivity in some samples and the complex angular dependent behavior at high fields and temperatures in Fig. 2 may be caused by the same reason. As addressed above, the state near the onset transition temperature is very complex with many kinds of contributions to the resistivity. Under these circumstances, both vortex motion and quasiparticle scattering will contribute to resistivity, which may result in a strange distribution of the conducting current and thus shows the strange resistance behavior.

Fig. 4 shows angle-dependent resistivity data measured at 1.9 K and 0.5 T on the three samples. As determined above, $\theta^{min}$ for sample 1, 2, and 3 are 176.3°, 176.7°, and 93.9° and are indicated by the red lines in Fig. 4. The corresponding crystal structures on the Se-Se plane is also shown to indicate the angular relationship between resistivity anisotropy and the crystal axis direction of the Se-Se plane.

**Discussion**

With the six-fold crystallographic structure, it is a surprise to observe a two-fold symmetry of resistivity in the superconducting transition region. One may argue that the two-fold symmetric feature in angle-dependent resistivity data is the possible ordering of the intercalated Sr atoms in a two-fold symmetry, like the strip structures. If those stripes have the same orientation, an in-plane rotational symmetry breaking may be induced. However, we have already studied the cleaved surfaces of $Sr_xBi_2Se_3$ single crystals by scanning tunneling microscope and no any regular patterns of intercalated Sr have been observed[20]. The random distribution of Sr atoms cannot break the crystal symmetry. Since the two-fold symmetry only emerges in the superconducting transition range, it is obvious that such feature is strongly correlated with the symmetry of the superconducting gap.

Intuitively, the two-fold symmetric resistivity curves imply that the superconducting gap may have a two-fold symmetric feature by considering the relation of the $H_{c2}$ and the gap value. According to the Pippard equation, we have

$$\xi_0 = \frac{\hbar v_F}{\pi \Delta(0)}, \tag{2}$$

here $v_F$ is the Fermi velocity, and $\Delta(0)$ is the superconducting energy at zero temperature. And by definition for a type-II superconductor, we have $H_{c2}(0) = \frac{\Phi_0}{2\pi \xi_0^2}$, where $\Phi_0$ is the magnetic flux quantum. Thus we get

$$H_{c2}(0) = \frac{\Phi_0 \pi}{2\hbar^2 v_F^2} \Delta(0)^2. \tag{3}$$

Now it becomes clear that the upper critical field will correlate with the gap, and the angle dependence of the c-axis resistivity with a two-fold symmetry can get a good explanation. Along $\theta^{min}$ directions, $H_{c2}$ reaches its maximum as shown in Fig. 3. One can derive that in the same direction, the superconducting gap value also has the maximum. Therefore, the red lines in Fig. 4 also indicate the maximum directions of the superconducting gaps. Our previous STS studies suggest that the superconducting gap of $Sr_xBi_2Se_3$ is nodeless but has an anisotropic component[20], which is qualitatively consistent with the existence of superconducting gap with a two-fold symmetry.

For sample 1 and 2, the angles of the gap maximum direction ($\theta^{min}$) approximately equal to 180° (0°) with the uncertainties less than 4°. Considering the inevitable experimental errors in the angle determination, such an uncertainty is acceptable. The direction of the maximum superconducting gap is most likely to be pinned along the Se-Se crystallographic axis as indicated by the lattice structure illustration in Fig. 4a, b. However, for sample 3 shown in Fig. 4c, the gap maximum direction is approximately along the direction which is vertical to the Se-Se crystallographic axis, also with the uncertainty less than 4°. Above all, we can conclude that the superconductivity in $Sr_xBi_2Se_3$ is nodeless and two-fold symmetric, and the gap maximum direction prefers to be parallel or perpendicular to the crystallographic axis. This is consistent with the earlier experiments by NMR[21] and angle resolved magnetocaloric[22] measurements.

According to the theoretical studies[23], for the crystal structure of $Cu_xBi_2Se_3$ and $Sr_xBi_2Se_3$, there are six irreducible representations characterizing the superconducting pairing symmetry, i.e., $A_{1g}$, $A_{1u}$, $A_{2u}$, $A_{2g}$, $E_u$, and $E_g$. Among them, only the $E_u$ representation is satisfied with our results, and the NMR[21] and angle resolved magnetocaloric measurements[22]. Since the $E_u$ represents the odd parity of Cooper pairing, our results here suggest that $Sr_xBi_2Se_3$ is a topological superconductor.

In summary, we performed the angle-dependent resistivity measurements on four $Sr_xBi_2Se_3$ samples with the Corbino-shape like electrode configuration. Dramatic two-fold symmetry features are observed in all the experiments at low magnetic fields and temperatures. This along with the $H_{c2}$ measurements and the former STS experiment provide the evidence of nodeless but two-fold symmetric superconducting gap. This is consistent with the prediction of topologically non trivial superconductivity in $Sr_xBi_2Se_3$. We also find that the gap maximum direction prefers to be pinned parallel or perpendicular to the crystallographic axis.

Note added: When preparing the manuscript, we became aware of another investigation which also achieves the conclusion about the two-fold anisotropy in the upper critical field measurements of superconducting Sr$_x$Bi$_2$Se$_3$ crystals[24]. However, considering that the current in their experiment is fixed in the plane, and the direction of the magnetic field is rotated in-plane, there will be an inevitable two-fold angle dependence of the Lorentz force contribution to the vortex motion. A two-point measurement with the current along c-axis was supplemented later by the same group and the results are consistent with ours here.

## Methods

**Sample growth.** We used the melt-growth method to grow the crystals. The stoichiometric amounts of Sr, Bi, and Se were sealed in an ampoule that was sealed within another ampoule (double-sealed) under 0.2 bar Ar, then the ampoule was placed horizontally in a small box furnace, heated up to 900°C, rocked to mix the liquid for 10 hours, cooled to 640°C at 0.5°C/h, and then quenched in liquid nitrogen. The inner ampoules were 2mm thick, had 10mm inner diameter, and were 15 cm long; the outer ampoules had similar dimensions and were 1mm thick. Typically, Sr$_x$Bi$_2$Se$_3$ was made this way in 30g batches.

**Resistive measurements.** The dc magnetic susceptibility was measured using a commercial superconducting quantum interference device with the vibrating sample magnetometer facility (SQUID-VSM, Quantum Design). The Laue x-ray crystal alignment system (Photonic Science Ltd.) was used to investigate the crystal orientation of the cleaved surface. The angle-dependent measurements are carried out on a physical property measurement system (PPMS, Quantum Design). The sample was cleaved along the Van der Waals layers and cut into square. The four-electrode method is performed and the electrodes are made on the cleaved surfaces of the sample with the Corbino-shape like configuration.

To eliminate the influence of the slight Hall signals on the raw data of angular dependence of resistivity, the resistivity taken at every angle has been averaged with positive and negative magnetic fields.

**Acknowledgements** We thank Guoqing Zheng and Liang Fu for helpful discussions. This work is supported by national natural science foundation of China (NSFC) with the projects: A0402/11534005, A0402/11190023; the Ministry of Science and Technology of China (Grant No. 2016YFA0300404, 2011CBA00100, 2012CB821403) and PAPD. The work in Brookhaven is supported by the Office of Science, U.S. Department of Energy under Contract No. DE-SC0012704. J.S. and R.D.Z. are supported by the Center for Emergent Superconductivity, an Energy Frontier Research Center funded by the U.S. Department of Energy, Office of Science.

**Author Contributions**

The samples were made by J.S., R.D.Z. and G.D.G. The angular dependent resistivity measurements and the Laue diffraction were done and analyzed by G.D. and helped by Y.F.L and H.Y. The magnetic properties were characterized by G.D. and H.Y. The manuscript was written by G.D. and H.-H.W. and supplemented by H.Y. H-H.W. coordinated the whole work and designed the configuration of the resistivity measurements with an in-plane magnetic field.

**Author Information** The authors declare no competing financial interests. Correspondence and requests for materials should be addressed to H-H.W (hhwen@nju.edu.cn).

# Figures and legends

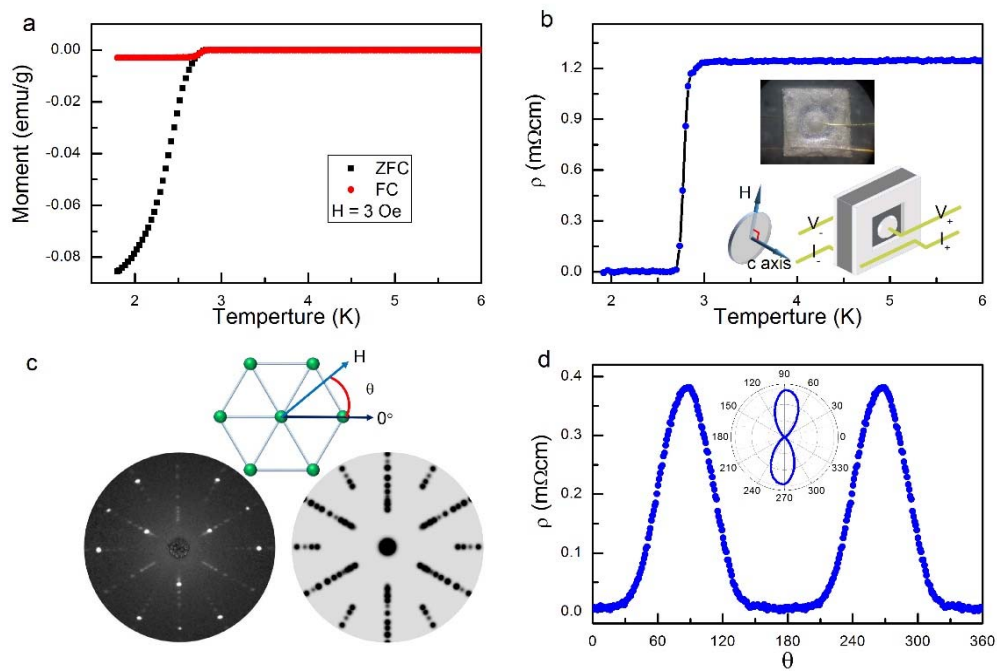

**Figure 1 | Superconducting properties of sample 1.** (**a**) Temperature dependence of magnetic susceptibility measured with ZFC and FC processes at an applied field of 3 Oe. (**b**) Temperature dependence of resistivity at zero field. The insets show the photograph of sample 1 with the electrodes (above) and the Corbino-shape like electrode configuration (below). The magnetic field is applied perpendicular to the *c*-axis, namely, within the plane of the sample. (**c**) The experimental Laue diffraction patterns of the cleaved top surface of sample 1 (left), theoretically simulated Laue diffraction patterns (right), and the derived lattice structure of the Se terminated surface (above). *θ* is the angle defined between the magnetic field and one selected crystal axis in the hexagonal plane. (**d**) The angular dependence of c-axis resistivity measured at 0.5 T and 1.9K. The inset shows the representation of the same set of data by polar coordinates.

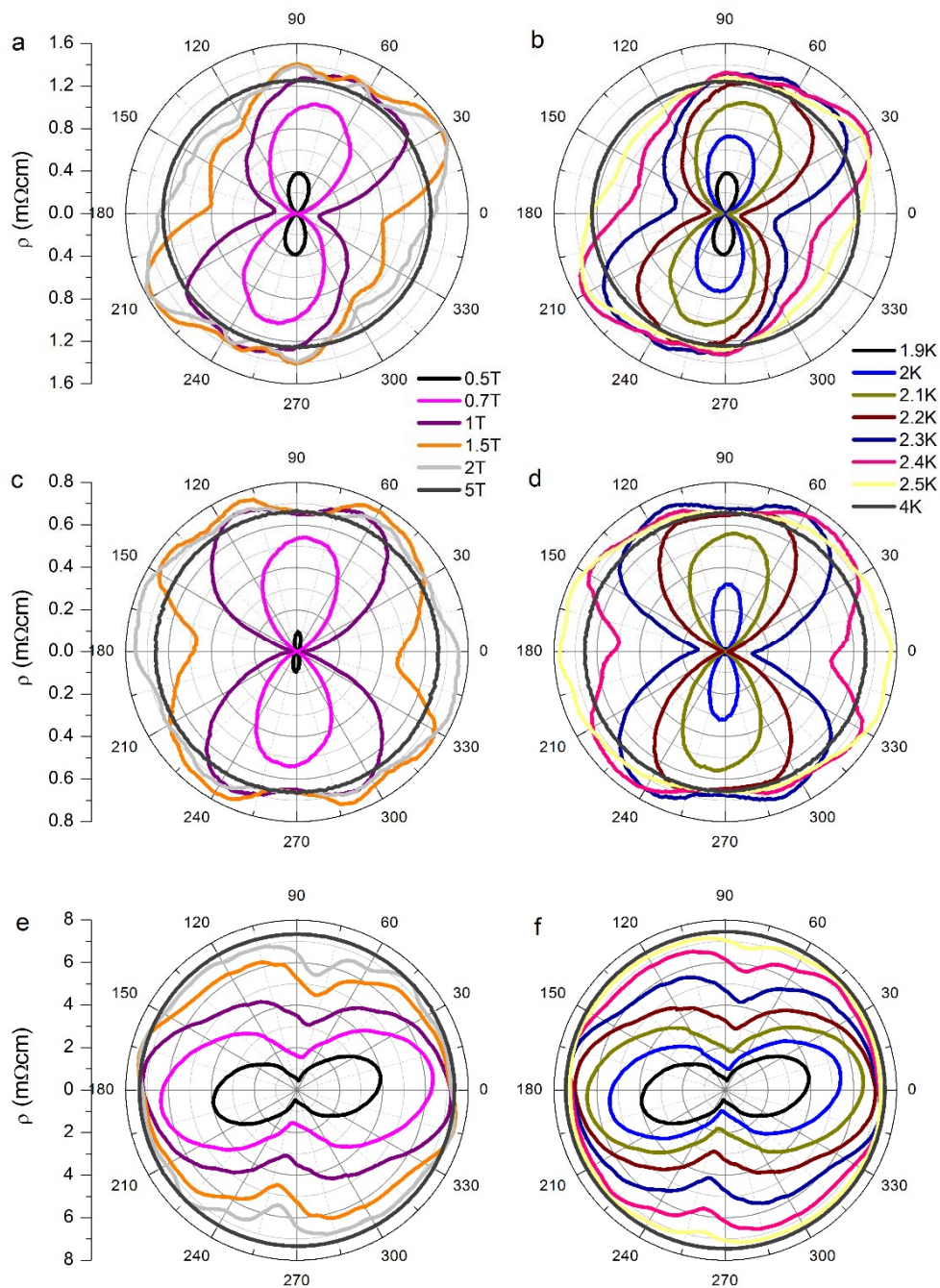

**Figure 2 | The angular dependence of resistivity of sample 1-3 measured at various fields and temperatures.** (**a**, **c**, **e**) Resistivity data measured at 1.9K and different magnetic fields for sample 1(in **a**), 2(in **c**), and 3(in **e**). (**b**, **d**, **f**) Resistivity data measured at 0.5 T and different temperatures for sample 1(in **b**), 2(in **d**), and 3(in **f**).

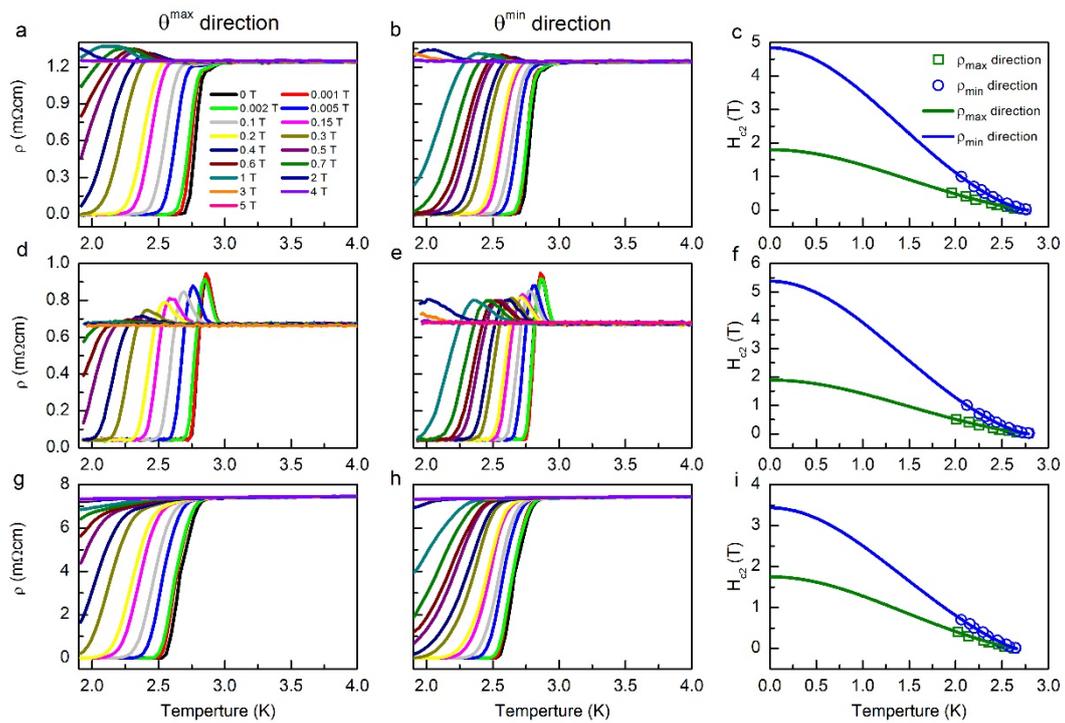

**Figure 3 | Superconducting transition at different magnetic fields and $H_{c2}$ anisotropy.** (**a**),(**b**) Data of sample 1 measured at $\theta^{max}$ and $\theta^{min}$. (**c**) Extracted upper critical fields $H_{c2}(T)$ for the two field orientations. The dots represent the experimental data and solid lines are the fitting results. (**d**-**i**) The similar results for sample 2 (**d**-**f**) and sample 3 (**g**-**i**).

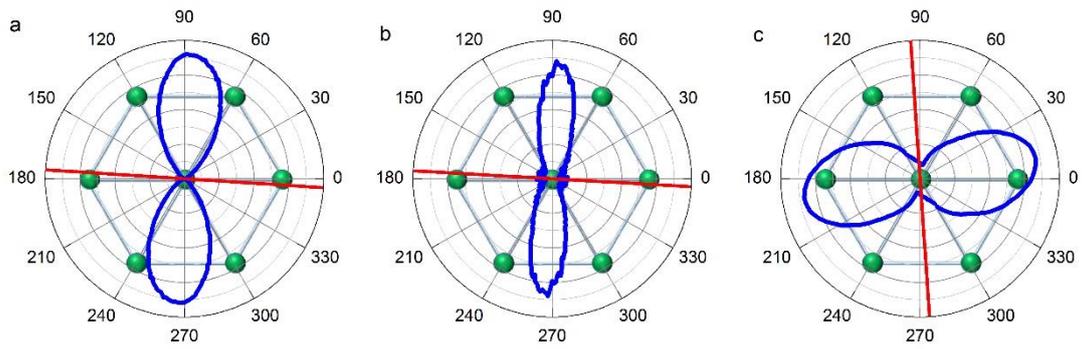

**Figure 4 | Analysis of orientation of the two-fold resistivity.** The blue curves in (**a**-**c**) are the angular dependence of the resistivity data for sample 1-3 respectively, measured at 1.9 K and 0.5 T. The red lines indicate $\theta^{min}$ direction for each sample. $\theta^{min}$ = 176.3° or −3.7° (in **a**), 176.7° or −3.3° (in **b**), and 93.9° or −86.1° (in **c**). The illustrations of the corresponding crystal structures are also overlaid below the data in (**a**-**c**). The green balls here represent the Se atoms on the terminated layer.

# Supplementary Note for the manuscript entitled Superconductivity with two-fold symmetry in topological superconductor $Sr_xBi_2Se_3$

**Supplementary Note 1**

In total, we have successfully measured four samples, all show the 2-fold behavior of resistivity. For samples 1-3, both the resistivity and Laue diffraction were successfully measured, while for sample #4, we did not manage to conduct the Laue diffraction measurement. Here we present also the data of sample 4. But the angle here does not reflect any relation with the crystallographic axes.

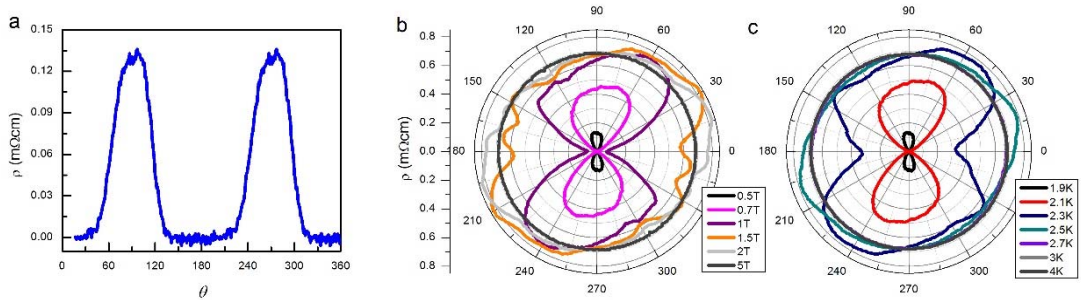

**Figure S1 | Angular dependence of resistivity of sample 4.** (**a**) The angular dependence of resistivity measured at 1.9 K and 0.5 T. (**b**) Data measured at 1.9 K at different magnetic fields. (**c**) Data measured at 0.5 T at different temperatures. All the data behave in the same way of sample 1, 2, 3. The two-fold symmetric feature is distinct in (**a**). The curves below 1 T in (**b**) and below 2.2 K in (**c**) have the simple dumbbell-shape. When at 5 T in (**b**) or 4 K in (**c**), the sample is in the normal state and the data are in the form of isotropic circles. Lacking of the Laue diffraction measurement of sample 4, the angle values in (**a**-**c**) have no relationship with the crystal axis in the hexagonal plane.

## Supplementary Note 2

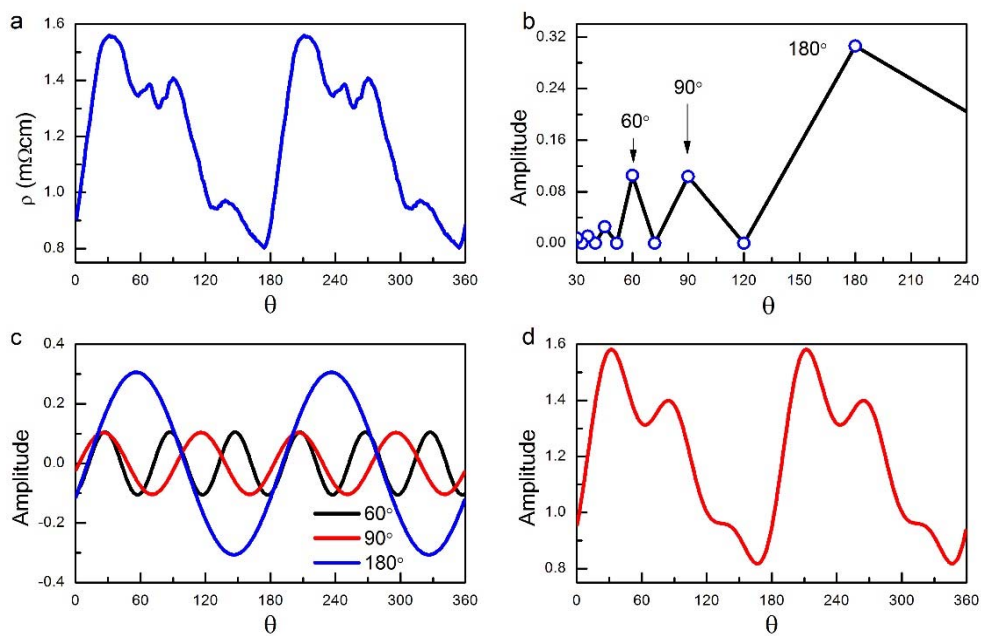

**Figure S2 | The Fourier transformation analysis on ρ(θ) data measured on sample 1 at 1.9 K and 1.5 T.** (**a**) The experimental data of sample 1. (**b**) The amplitude values of all the components obtained from the Fourier transformation analysis. One can find that there are three main period components in the experimental data, i.e. 60°, 90°, 180°. (**c**) The waveforms of the 60°, 90°, 180° components. (**d**) The curve synthesized by the accumulation of the three waveforms in (**c**) and a constant shift of 1.19 without any periods is added to compare with the experimental data. The synthesized curve is consistent with the raw data in (**a**) implying that the three components of 60°, 90°, and 180° represent the key features of the experimental data. The 90° period component may be the result of the frequency doubling of the two-fold symmetric 180° period in the process of Fourier transformation. The 60° period component may contain the information contributed by the six-fold crystal structure.

## Supplementary Table 1

By the fitting in Fig. 3c, 3f, 3i, we get:

|  | $\theta^{min}$ direction | | $\theta^{max}$ direction | | anisotropy |
| --- | --- | --- | --- | --- | --- |
|  | $\mu_0 H_{c2}$ (T) | $\alpha$ | $\mu_0 H_{c2}$ (T) | $\alpha$ | $H_{c2}^{min} / H_{c2}^{max}$ |
| Sample 1 | 4.844±0.015 | 1.44±0.04 | 1.801±0.0057 | 1.25±0.04 | 2.69 |
| Sample 2 | 5.380±0.034 | 1.41±0.04 | 1.8918±0.00007 | 1.24±0.04 | 2.84 |
| Sample 3 | 3.437±0.084 | 1.20±0.046 | 1.752±0.0058 | 1.19±0.05 | 1.96 |

The anisotropy is defined by $H_{c2}^{min} / H_{c2}^{max}$. $H_{c2}^{min}$ is $H_{c2}$ at the $\theta^{min}$ direction and $H_{c2}^{max}$ is $H_{c2}$ at the $\theta^{max}$ direction.